# Concepts and Algorithms for Agent-based Decentralized and Integrated Scheduling of Production and Auxiliary Processes

**Felix Gehlhoff[1], Alexander Fay[1] (Senior Member, IEEE)**
[1]Helmut Schmidt University / University of the Federal Armed Forces Hamburg, Germany

Corresponding author: Felix Gehlhoff (e-mail: felix.gehlhoff@hsu-hh.de).

This work has been supported by the German Federal Ministry of Economy and Energy in the frame of the ZIM program under grant no. FKZ 16KN069722. All responsibility for the content remains with the authors.

**ABSTRACT** Individualized products and shorter product life cycles have driven companies to rethink traditional mass production. New concepts like Industry 4.0 foster the advent of decentralized production control and distribution of information. A promising technology for realizing such scenarios are Multi-agent systems. This contribution analyses the requirements for an agent-based decentralized and integrated scheduling approach. Part of the requirements is to develop a linearly scaling communication architecture, as the communication between the agents is a major driver of the scheduling execution time. The approach schedules production, transportation, buffering and shared resource operations such as tools in an integrated manner to account for interdependencies between them. Part of the logistics requirements reflect constraints for large workpieces such as buffer scarcity. The approach aims at providing a general solution that is also applicable to large system sizes that, for example, can be found in production networks with multiple companies. Further, it is applicable for different kinds of factory organization (flow shop, job shop etc.). The approach is explained using an example based on industrial requirements. Experiments have been conducted to evaluate the scheduling execution time. The results show the approach's linear scaling behavior. Also, analyses of the concurrent negotiation ability are conducted.

## I. INTRODUCTION

Running an industrial production facility is, and has always been, a challenging task. Production companies strife for an optimized production while satisfying their customers. However, customer needs and the ways to satisfy the latter have changed. Nowadays, it is not enough to simply maximize the output of a manufacturing plant. Rather, the customers demand high-quality products with short life cycles that are also highly customized [1, 2]. Different initiatives all over the world try to develop solutions and conceptual frameworks that enable companies to deal with this new reality.

There is, for example, the Smart Manufacturing Leadership Coalition, which represents many of the major North American industrial companies and universities. Their goal is to provide more flexible and resource-efficient manufacturing by integrating the 'voice, demands and intelligence' of the customer into the manufacturing process [3]. Among the top priorities they identified range the development of tools and software that enable automated decision-making as well as the establishment of data protocols and communication standards.

The vision of Industry 4.0 (I4.0) proclaims the automated decision-making within the digital representations of products and resources, such as that products control their own manufacturing processes [4]. The German initiative of the Industry 4.0 platform provides reference scenarios and solution concepts for this vision. There is, among others, the scenario of the Order Controlled Production that points out the challenges that traditional centrally controlled and highly specialized machinery face when being confronted with changing customer requirements and disruptions on the production floor. The Industry 4.0 platform emphasizes the need for standardized communication and coordination mechanisms that enable companies to engage in inter-company negotiations, which facilitates the optimal utilization of their resources and increases their flexibility [5].

Interestingly, the challenge of rising complexity and possible solution techniques have been known for a long time. For instance, the reference model for Computer Integrated Manufacturing (CIM) already formulated requirements to their model such as modularity, standardized semantics, interoperability, or simple expandability [6]. However, the fact that these topics are still addressed by today's initiatives 30 years later indicates how difficult it is to develop generic and universally applicable solution concepts while dealing with constantly changing information and manufacturing technologies. In addition, not only has the speed of manufacturing and information technology development increased [7] but the latter has also facilitated a whole new set of possibilities. Customer interfaces and interactions such as e-commerce, real-time order configuration, and live production monitoring [8] are becoming increasingly important for customers.

Companies have to react by frequently changing their product portfolio and, thus, changing their resources as these are mostly designed for producing a small variety of products in large quantities [9]. These traditional mass production systems, i.e. production lines, are often directly coupled, for example by conveyor belts. In a production line there is usually no need to plan or schedule such transport processes and related buffer processes for the workpieces as they usually are fully predictable and don't pose capacity constraints on the system (i.e. they are usually not the bottleneck). However, their rigid configuration and limited flexibility when it comes to changing products and layouts hinders flexible transport processes between different resources [10]. The need for more flexible transportation solutions also arises for large workpieces, such as rotor blades for wind turbines, that cannot be transported by conveyors. When breaking up these rigidly connected transportation systems, the number of active transportation resources increases and can pose additional requirements such as heterogeneous capabilities, individual schedules, possible collisions or further shared resources (SRs) such as operators to control the transportation resources. In the following, buffer, transportation and shared resources are subsumed under the term auxiliary resources (ARs) and auxiliary processes (APs) respectively. Because of these additional requirements, the transportation processes can become the bottleneck of the production facility. In addition, the number of possible sequences through the system increases significantly with each newly added resource. Thus, a process planning approach is required that determines and evaluates possible sequences of operations on resources. An operation refers to a specific process where a workpiece is manipulated by a resource [11]. These factors are also important to determine the quality and costs of specific resource combinations [12]. The output of such a process planning run is a process plan that indicates possible machines for each operation, which the workpiece requires during production. An approach to solve the outlined problem of decentralized process planning has been proposed by the authors in [12]. Therefore, this paper focuses on the process scheduling part of the problem.

The question arises if central control methods can be used for the more flexible production and transportation systems. Centralized systems have several drawbacks when being applied to complex problems in general. These drawbacks of centralized and hierarchical systems include expensive and time-consuming reconfigurations, a single computer as the potential bottleneck and single-point-of-failure, flows of information across hierarchies that increase latency time of decision-making, and a tendency to struggle with real-time events [13]. Thus, they do not scale well with growing system size and increasing communication needs, and flexible production approaches are difficult to realize [14]. Therefore, more flexible and scalable approaches and a respective communication architecture are needed [15].

Two needs arise from this analysis: First, a scheduling approach is needed that can handle the rising complexity and flexibility requirements while integrating the different resources into a coherent schedule [1]. Thus, it needs to avoid delays which can arise due to insufficient coordination of the resources. Second, these functionalities must be implemented within a decentralized architecture to avoid the already mentioned drawbacks of centralized solutions.

Making use of decentralized intelligence within products and resources while equipping them with ubiquitous communication technology is a promising approach to deal with the rising complexity within manufacturing systems [16]. One solution technology that has been identified as suitable for implementing decentralized intelligence is Multi-agent systems (MAS). MASs are especially suitable for solving complex problems due to the possibility of decomposing the problem and solving it within a decentralized architecture, their flexible communication structure, their capability of parallel processing due to autonomous entities, and higher robustness because an MAS does not usually have a single-point-of-failure. [17, 18]

However, even MASs are limited to some extend when the number of exchanged messages increases. Berna-Koes et al. conclude that, due to the overhead of agent communication languages and the considerable processing time of ASCII-coded messages, the number of messages that is exchanged between the agents constrains the systems size and performance [19].

Considering the above-stated needs for decentralization and integrated scheduling as well as the rising complexity and flexibility requirements, the following research goal is proposed:

**Development of a decentralized and scalable agent-based architecture and respective algorithms that integrate production and ARs during scheduling.**

The remainder of this contribution is structured as follows: First, the detailed requirements for the architecture and algorithms are elaborated in Section II. This is followed by a literature review in Section III that analyses different scheduling approaches and architectures for solving similar problems. Section IV gives a brief overview of the approach to achieve the proposed research goal. Section V describes and evaluates an agent-based architecture and the negotiation protocol and provides details regarding the exchanged messages. Section VI presents the application example and the fundamentals that are utilized in Section VII to describe the details of the coordination mechanisms. This section is followed by an evaluation in Section VIII. Section IX closes the article with a conclusion and an outlook.

## II. PROBLEM DESCRIPTION AND REQUIREMENTS

This contribution advances the research in the field of decentralized scheduling algorithms, based on MASs, for integrated production and AR scheduling. Most of the specific requirements that will be outlined in the following stem from an industrial project within the domain of wind turbine rotor blades manufacturing. However, as outlined in Section I, there is a general necessity to develop concepts for decentralized scheduling architectures that can, for example, also be applied to scenarios such as large-scale manufacturing networks [20]. Thus, even though the application example is very specific, the detailed analysis facilitates the understanding and the applicability of the approach to other problem areas. In the following, this section describes the different requirements that the approach has to fulfill and their origins. Please be referred to a full and mathematical description of the problem in [21] as well as a more detailed description of scheduling fundamentals in [11].

The most important factor behind the need for new scheduling and control solutions is the rising system complexity [1]. System and organizational complexity look at interactions between system components and how systems can be organized [22]. This is also the focus of this approach, i.e. the process and structure of the interactions between the agents that represent the system components. Following the complexity characterization of [23] there are three different categories to system complexity:

- First, there is organizational complexity which refers to **process** and **structural** complexity. It describes the number and diverseness of process flows and system elements respectively.
- Second, there is time-related complexity that is divided into **static** or **dynamic** properties of a system. Dynamic complexity refers to "number and diverseness of process-flows, systems elements, their relations and properties in time dependent course" [23] whereas static complexity focusses on the status of the system at a concrete point in time.
- The systemic complexity describes the third category and deals with **internal** and **external** complexity referring to inside and outside connections of the system. Thus, the entities of the system, i.e. those processes, system elements etc. that lie within the system boundary, determine the internal complexity. External complexity refers to those elements that lie outside the system boundary.

In the following analysis these dimensions are combined and addressed in different sub-sections. The only category that is out of scope for this contribution is the structural static external complexity that refers to, for example, the number of suppliers involved. As the system border lies at the production facility with incoming orders as the interface, additional supply chain related connections are not considered. In the following, the requirements (Req) will be derived and categorized according to the complexity dimensions.

### A. Structural Complexity

This sub-section analyses the structural aspects of the requirements, which comprise the architectural properties of the system (static, internal), the dynamic properties of the system's components (dynamic, internal) and the order-related requirements (dynamic, external).

**Req 1 (Scalability)**
**1a (Scalable Communication Architecture):** The approach needs to be able to cope with a large number of resources in the coordination process. Thus, the execution time of a scheduling run, which schedules all operations of an order, must not grow exponentially with growing system size but only linearly. To limit the number of exchanged messages, which are a major driver of execution time within MASs, the approach must provide a scalable communication architecture for the participants [24].

**Req 2 (Decentralization):** There must be no central control authority as the system shall be organized in a decentralized way due to the reasons given in Section I. A further reason to decentralize functionalities is often the organizational setup: a heterogeneous production network that has no central authority [20]. Another reason is the need for encapsulation, for example due to restrictions to share information about underlying capability models.

**Req 3 (Dynamic Events)**
**3a (Runtime Changes of Resources):** The resources can change during the runtime of the system, i.e. there can be new resources joining or old resources leaving the system at any point in time [20].
**3b (Disturbance Management):** The feasibility and quality of a specific operation sequence depends on dynamic factors (such as failures) of the resources. Thus, the approach must be able to cope with the dynamic events, i.e. disturbances, on the shop floor [25].

**Req 4 (On-line Order Generation):** Orders are generated on-line, meaning that there can be new orders at any point in time [26].

### B. Process Complexity

This sub-section comprises requirements due to the different process flows to include in the approach and their respective static properties. Thus, the required scope of the scheduling processes is determined and their respective constraints. Also, the process requirements from incoming orders external are addressed.

Further, the dynamic process complexity, which refers to the behavior of the different process flows, is analyzed. This refers to the desired results of the scheduling process and the dynamic properties of the scheduling processes during runtime.

**Req 1 (Scalability)**
**1b (Short Execution Time):** The approach shall be practically applicable for large scale applications. It must provide a scalable architecture (Req 1a) and exhibit a sufficiently short execution time for scheduling runs for realistic scenarios. To determine a suitable threshold, the study conducted by *Nah* is utilized. There, the tolerance of users to wait for information on the web lies at around two seconds [27]. This is in line with findings of Stadnik & Nowak. They concluded a significant increase in the bounce rate, i.e. the rate of customers leaving a page before it has loaded, with increasing page load times [28]. Thus, imagining a user-interface where a customer or operator is awaiting an indication of the estimated finish date and price of an order, the scheduling execution time for an order should not exceed two seconds. This is also in line with the industrial Req from our industry partner. The execution time is posed by a user outside the system boundary (i.e. external) and addresses dynamic process properties of the system. In addition, a short execution time is necessary to provide solutions for dynamic events (Req 3).

**Req 5 (Scheduling Scope):** The scheduling procedure must be able to incorporate production and AR scheduling. SRs refer to, for example, operators or tools that are needed by multiple production resources to fulfill an operation [29] or a shared physical space such as a single crane rail for the movement of several overhead cranes [21]. As further outlined in [11], these resources are addressed on different *stages* of the scheduling process. Each type of resource that needs to be scheduled for a production process constitutes an additional stage of the scheduling problem. The number of necessary stages to include in the scheduling process as well as the number of resources on each stage depend on the application. For example, after the scheduling of production resources on the first stage there can be ARs to be considered on subsequent stages for each scheduling run.

**Req 6 (Process Planning):**
Furthermore, an approach for decentralized process planning is required. Often, the interdependencies between logistics and production processes are neglected during process planning [30]. However, there can be combinations of production processes that are, due to the lack of suitable transportation means between them, undesirable, which requires the incorporation of this dependency during process planning. Even though the process planning part of the approach is addressed in a different publication [12], this Req is still mentioned here to enable a sufficiently detailed analysis of the state of the art.

**Req 7 (Logistics Constraints)**
**7a (Buffering and Handling):** The handling of large workpieces requires special constraints to be met for generating feasible plans. These include limited buffer capacities, as each workpiece requires a dedicated place to be stored at (if there is one at all) and, for the same reasons, a transportation resource cannot unload and load a workpiece in the same run [31]. In addition, handling abilities of transportation resources need to be accounted for [30].

**7b (Blocking Constraint):** The difficult buffering process of large workpieces also results in a strong dependency between production and transportation processes. As long as a workpiece has not been picked up from the production resource, the latter cannot start with a new operation as no second workpiece can be delivered at the same time. Also, a finished job cannot leave the machine if no buffer is available and the next machine is not ready yet or the workpiece is not picked up by a transportation resource. This so-called blocking constraint is defined by Mascis & Pacciarelli as the forced remainder of a job on a machine until the next (*production*) *machine* becomes available [31]. This will be adopted here so that the job remains on a production machine until picked up by a transport resource that delivers it either to a separate buffer place or to the next production machine.

**7c) (Collision Avoidance):** The movements of overhead cranes are strongly restricted by the other cranes on the shared crane rail as it is not possible for the cranes to move past each other. Zhang & Rose state that crane interference is one of the main factors that affect crane utilization [32]. The approach thus has to include the possibility to derive collision free plans within the decentralized setup, representing each SR by a dedicated autonomous entity.

**Req 8 (Variable Resources and Processes):** It must be possible to parametrize all processes and resources in the system individually. Thus, required processing times of each operation, setup times and capabilities, i.e. possible steps per machine, are not constrained by the approach. They exclusively "depend on the machine on which [the operation] is processed" [33].

**Req 9 (Variable Operation Sequences):** Customer orders should not be constrained by given operation sequences. Thus, the approach has to be applicable for any type of production environment, be it flow shop, job shop or open shop where "there are no restrictions with regard to the routing of each job through the machine environment" [33].

**Req 10 (Detailed Auxiliary Process Scheduling)** Besides the scope of the scheduling approach (Req 3) it is important to further clarify the level of scheduling granularity of the different processes to be scheduled.

**10a (Transportation Scheduling):** Factories dealing with large workpieces, such as wind turbine equipment or aircraft parts, usually rely on overhead cranes to execute necessary transportation processes. These processes require non-negligible amounts of time during loading and unloading of the workpieces. Thus, they influence the production operations, especially if the transport operations are a bottleneck [34]. The approach needs to incorporate load and unload times, setup times of transportation resources within scheduling, as well as logistics abilities, such as handover abilities, during process planning and scheduling. Furthermore, the level of integration of the transportation scheduling needs to be in line with production scheduling, meaning that the degree of detail reflects a dedicated resource with exact start and end times of the transportation operation. This excludes, for example, approaches where the duration of transportation processes is reflected by an increased duration of production time or a single transportation resource that does not adhere to any physical and timing constraints.

**10b (Detailed SR and Buffer Scheduling):** The integration level for SR and buffers during scheduling must also enable the distinction of each applicable resource for the exact duration of each process [21]. Hence, the respective resources actively manage their schedules [35]. This excludes approaches where, for example, the buffer is simply represented as an infinite queue or required tools are modelled as properties of a process (instead of dedicated entities with their own schedules).

**Req 11 (Schedule Stability and Optimization):** The approach must respect schedule stability. This is important in terms of worker preferences [36], as workers prefer predictable working conditions, and auxiliary material deliveries that would otherwise have to be re-planned and re-routed. Hence, changes to the already scheduled operations are undesirable in the context of this contribution. This might result in sub-optimal solutions because not all operations can be re-arranged in a scheduling run. However, it is important to utilize the possibility of scheduling new operations between already existing ones (in-between scheduling) to generate good solutions with regard to the throughput time of an order.

**Req 12 (internal, Concurrency):** The scheduling process itself must be stable, even in concurrent, i.e. simultaneously executing, scheduling runs. Within scheduling, this is especially important regarding the so-called 'indecision problem' [37] that refers to the interdependence of proposed timeslots to different orders. A timeslot (or a partially overlapping time interval) should not proposed to two entities at the same time. Otherwise, in case of multiple allocations of orders to these timeslots, rescheduling or restarting the scheduling process is required, which is not desirable as it significantly increases the execution time (see Req 14). The approach must also exhibit freedom of deadlocks during concurrent scheduling.

The following table summarizes the complexity dimensions of the different requirements of the approach.

TABLE I. Requirements with respect to the Complexity Dimensions

|  | Process Complexity | | Structural Complexity | |
| --- | --- | --- | --- | --- |
|  | Dynamic | Static | Dynamic | Static |
| Internal | 10,11,12 | 5,6,7,8 | 3 | 1a,2 |
| External | 1b | 9 | 4 | Not applicable |

The following section will analyze different approaches and their respective contribution to the requirements.

## III. LITERATURE REVIEW - ANALYSIS OF THE STATE-OF-THE-ART WITH RESPECT TO THE REQUIREMENTS

The problem of scheduling for flow shop, job shop or open shops is an NP-hard problem [38]. Thus, it cannot be optimally solved in acceptable computation time for realistic instances [39]. At the same time, the requirement of a decentralized approach makes it difficult to apply many commonly used (meta-) heuristic solutions to the problem, such as Genetic, Particle Swarm, or Tabu-Search algorithms as these are usually designed in a central way. The contributions that are presented in the following therefore concentrate on agent-based approaches. The different contributions were selected according to the following search criteria: Initially, relevant reviews in the field have been analyzed to find important approaches to include in the review ([1], [38], [40], [41]). Furthermore, a key word search on available databases has been conducted to determine further important contributions in terms of number of citations. Lastly, recently published contributions that tackle similar problems, i.e. address similar requirements, have also been added to the review. The following table summarizes the findings of the review and points out notable properties and features of the analyzed approaches as well as a brief summary of the fulfillment of the requirements.

**TABLE II.** Literature Review

| Contribution | Scheduling granularity | Architecture | Notable features | Notable differences | R1 | R2 | R3 | R4 | R5 | R6 | R7 | R8 | R9 | R10 | R11 | R12 |
|---|---|---|---|---|---|---|---|---|---|---|---|---|---|---|---|---|
| [42] | Scheduling | Hierarchical | Material provisioning and due dates are incorporated | Not included: transport, buffer & SRs, setup times, PP | ◐ | ● | ● | ● | ○ | ○ | ○ | ◐ | ◐ | ◐ | ◐ | ◐ |
| [35] | Scheduling | Hierarchical | Full integrated SR scheduling | Not included: transport & buffer resources, PP | ◐ | ◐ | ● | ● | ◐ | ○ | ○ | ◐ | ◐ | ◐ | ◐ | ◐ |
| [43] | Mixed | Heterarchical | Product types compete for order release | Transports not fully integrated, no setup times, no SRs | ◐ | ● | ○ | ◐ | ○ | ◐ | ○ | ○ | ● | ◐ | ○ | ◐ |
| [37,44] | Scheduling | Holonic | Backw. and forw. messaging, concurrency management by OA hosting restriction | No auxiliary process integration, higher communication intensity, unclear PP | ◐ | ● | ● | ● | ○ | ◐ | ○ | ● | ● | ● | ● | ● |
| [15] | Control | Decentralized | Differentiated RA types, state diagrams for the agents, material flow capabilities | No explicit scheduling, logistics constraints only partially fulfilled | ● | ● | ● | ● | ◐ | ○ | ○ | ◐ | ◐ | ○ | ○ | ◐ |
| [45–47] | Control | Decentralized | Desired processes modelled by physical properties, backw. messaging | Unlimited buffers in front of machines, concurrency issues not solved, no SRs | ◐ | ● | ● | ◐ | ● | ◐ | ● | ● | ● | ◐ | ◐ | ○ |
| [48,49] | Scheduling | Hybrid (hierarchical elements) | Incorporates preferences from higher levels via voting | Only the transp. costs are considered, no in-between scheduling, corrections required | ◐ | ◐ | ● | ● | ◐ | ○ | ○ | ◐ | ◐ | ◐ | ◐ | ○ |
| [50] | Scheduling & Dispatching | Hybrid | Integrated with RFID and simulations | Assumes unlimited buffers, no full transport integration, no SRs, no blocking constraint | ◐ | ◐ | ● | ● | ◐ | ○ | ○ | ◐ | ◐ | ○ | ◐ | ◐ |
| [51] | Scheduling | Holonic & Hybrid | Pheromone-based comm., coordination scheme changes (hierarchical vs. heterarchical) | No transport & SR scheduling, plan stability not addressed, no PP | ◐ | ◐ | ● | ● | ○ | ◐ | ○ | ● | ● | ○ | ◐ | ◐ |
| [52] | Scheduling & Dispatching | Holonic & Hybrid | AGV holons, system re-ornization | Transport orders are dispatched (not scheduled), no transp. details considered, no PP | ◐ | ◐ | ● | ● | ◐ | ○ | ◐ | ● | ● | ◐ | ○ | ◐ |
| [53] | Mixed | Decentralized | Sequential scheduling, flexible process plans | No transports, PP is not based on interface-oriented cap. Models, OA knows the machines a priori | ◐ | ● | ● | ● | ○ | ◐ | ○ | ● | ● | ○ | ○ | ◐ |
| [54] | Mixed | Hierarchical | Supervisor for stability, earliest start and latest finish concepts, currency concept | see [53] | ◐ | ◐ | ● | ● | ○ | ◐ | ○ | ● | ● | ● | ◐ | ◐ |
| [32] | Scheduling | Central | GA integrated with simulation, collision avoidance, detailed transport scheduling | All order and resource information are known upfront, centralized approach, no handling abilities | ◐ | ○ | ◐ | ○ | ○ | ○ | ● | ◐ | ◐ | ◐ | ○ | ○ |

| | | | | |
|---|---|---|---|---|
| ◐ | Partly fulfilled | ◐ | not fully disclosed but likely capable of | ● fulfilled |
| ○ | not fulfilled | ◐ | not fully disclosed but likely not capable of | |

The results of the review show that among the (completely or partially) decentralized approaches, there is a gap regarding the inclusion of all required functionalities (Req 5) and logistics constraints (Req 7). These problems have not yet been sufficiently solved within a decentralized architecture. The approach of Zhang & Rose provides valuable inputs for the required logistics-related concepts and the collision avoidance algorithms [32]. Also, it is the only approach that fully integrates transportation scheduling but does not address SR scheduling. Kovalenko et al. address important aspects for the decentralized process planning but also do not provide a sufficiently integrated solution with regard to auxiliary process scheduling and logistics constraints [45–47].

Concurrency is partially addressed by some approaches, but there is only one detailed solution provided in [37, 44]. Sousa & Ramos suggest to monitor the resources required by currently active orders to avoid overlapping sets of those resources. Hence, a new order is only introduced into the system if there is no potential conflict or overlap with the resource requirements of currently active orders. This approach might result in long delays in large systems where there are many resources required (and maybe required multiple times) by the active orders. The idea of a certain form of blocking mechanism might still be a possible solution to the problem and will be elaborated on in the subsequent sections.

Sousa & Ramos are also the only ones who present a scheduling approach that considers schedule stability (Req 11) [37, 44]. However, as the approach uses a resource-oriented coordination approach that increases communication load, it is still an open question how this Req can be addressed in a better scalable architecture. In [43], for example, the schedule stability is accounted for but the approach only schedules two orders in the future that are added at the end of the Resource Agent's (RA) schedule, which is a rather simple situation.

Regarding the scalable architecture and the required execution time, there is no scheduling solution that fulfills Req 1. Marschall et al., for example, develop a scalable architecture, but their approach does not explicitly schedule the operations but rather implements a control solution where the next operation is organized after the current operation has been finished [15]. Many of the other approaches exhibit an exponential increase in the number of messages [42], require corrections during scheduling [49] or do not provide sufficient details of their approach [53]. Approaches with central elements are also limited with regard to their scalability (Req 1) [32, 51]. A more detailed analysis of different architectures and their scalability is provided by the authors in [11], concluding that heterarchical and hybrid architectures in general exhibit quadratic growth of number of messages, whereas mediator-based architectures can be realized with linear growth, but violate the decentralization Req (Req 2).

As there is no approach or architecture that fulfills all requirements and some requirements have not been sufficiently solved by any of the analyzed approaches, the postulated research goal requires new research artifacts, i.e. architectural concepts and parameters as well as algorithms. From the identified gaps and the outlined focus of this contribution on the scheduling part of the problem, two main research questions (RQ) can be derived:

**RQ 1:** What features requires an architecture of an MAS-based scheduling system to exhibit a linear growth of the number of exchanged messages between the agents during a multi-stage coordination process?
**RQ 2:** How can the problem of integrated production and AR scheduling, based on flexible process plans, be solved within a decentralized architecture?

RQ 1 focuses on the architectural design of the approach whereas RQ2 addresses the algorithmic parts. A combination of both parts is required to fulfill the postulated requirements. To answer the RQs, an approach composed of different parts has been derived and is outlined in the following section.

## IV. OVERVIEW OF THE APPROACH

The approach presented in this contribution consists of different parts that each address a part of the outlined RQs (see Fig. 1). RQ1 is addressed in the Scheduling Architecture where questions like protocol design, deadlock and reiterations avoidance and required information within the exchanged messages are tackled. RQ2 focuses on the required algorithms for the integrated scheduling approach in the Scheduling Algorithms.

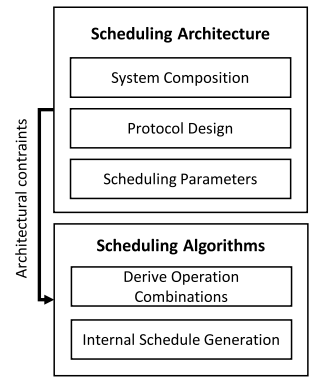

**FIGURE 1.** Conceptual Overview.

This contribution focuses on the scheduling architecture and algorithms of the approach. It advances the fundamentals described in ([11, 21]). The problem of collision avoidance and details about the aspects of SR scheduling in general will be addressed in future publications. As already mentioned, Zhang & Rose [32] provide promising fundamentals for collision avoidance whereas the approach in [35] presents an interesting solution for the problem of SR scheduling. However, even though these approaches are good solutions to the related aspects of the overall problem, they need to be integrated into a unified architecture.

Also, relevant aspects of disturbance management, which are not the focus of this contribution but often pose important requirements for practical applications, have been described by the authors already in [55] and [56]. Fig. 1 outlines the different parts and their interdependencies. Generally speaking, the scheduling architecture (functional composition, decentralized protocol design and the information model) create the solution space for the process planning algorithm and the scheduling algorithms. Hence, the algorithms need to incorporate the given functional decomposition, protocol and message structure and the information model.

## V. SCHEDULING ARCHITECTURE

This section explains the architecture (based on and extending prior work [21]) that incorporates details on the system's components and their static relationships, the components' functionalities as well as their dynamic interactions. It also describes some important aspects concerning the exchanged messages during these interactions and finally evaluates the architecture with respect to the postulated requirements.

The proposed MAS architecture consists of different agents that each fulfill their respective role in the system and additional components such as databases or webservices. Here, the approach follows well-known architectures that propose different agents (or holons) for resources, orders, products (or related information) and services [51, 57, 58]. The following figure displays the different types of agents that are not already part of FIPA standardized frameworks such as JADE [59]. Hence, the Agent Management Service (AMS) Agent that manages the agents on the platform (e.g. starting them) and Directory Facilitator (DF) Agent, which keeps a register of the capabilities of the agents, the agent for the graphical user interface and further auxiliary agents for debugging are part of the proposed architecture but omitted from the figure.

The functions fulfilled by the agents are connected to their type and name. The Order Agent (OA) is responsible for hosting the respective Workpiece Agents that are required for the order, for monitoring the scheduling process of each workpiece and relating the order fulfillment to an Interface Agent to implement the schedule. Note that the creation of separate Workpiece Agents is not necessary for every application. This is only sensible if the workpieces are distinguishable and handled separately on the production floor, i.e. it is not necessary for small pieces that are produced within one lot or batch. Therefore, to facilitate the reading process of this contribution, only the term OA is used instead of distinguishing between OA and Workpiece Agent. The

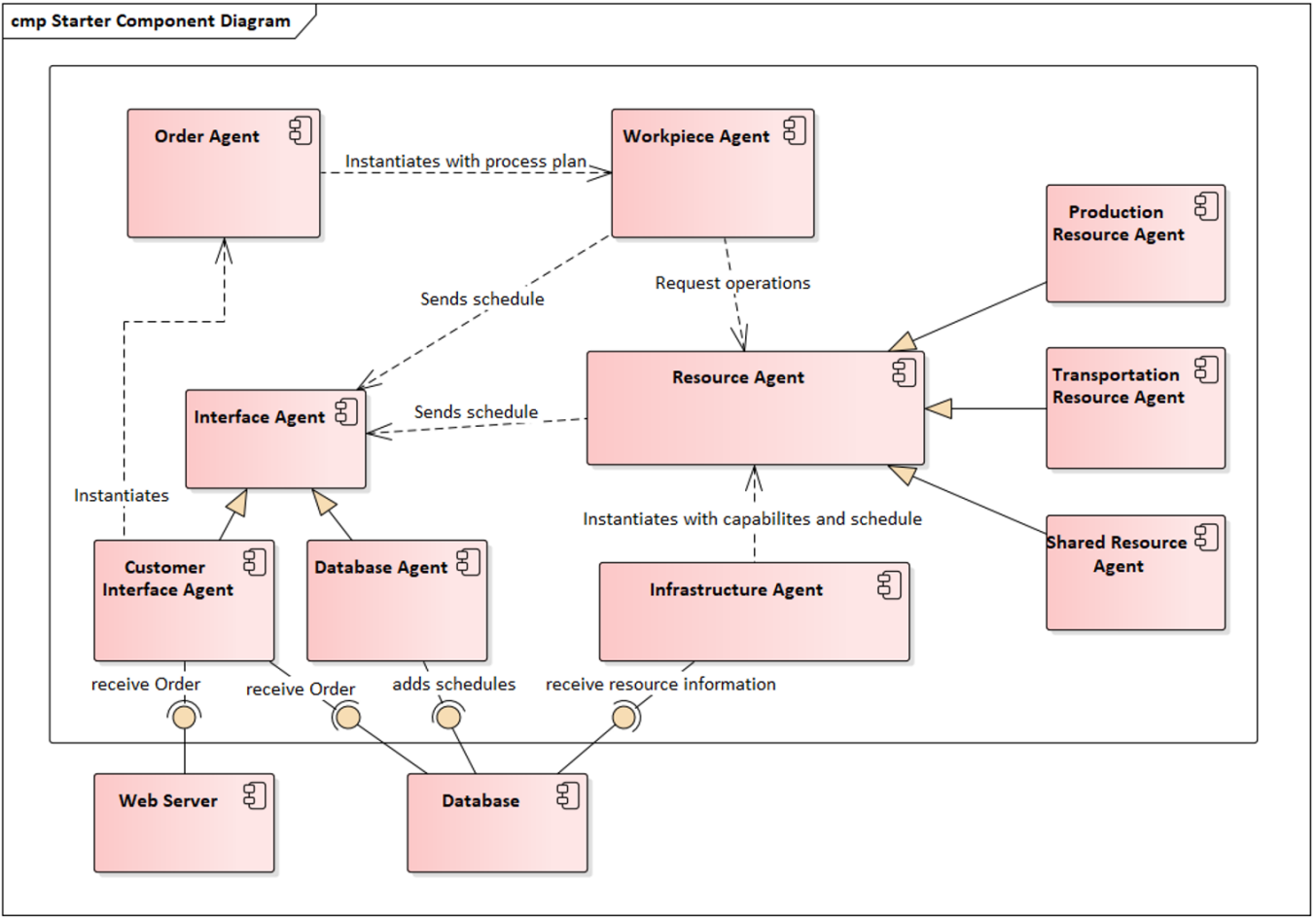

**FIGURE 2.** System composition with agent types.

Resource Agents (RAs) manage their resource's schedule and, for the process planning part, their resource's capability model (see [12]). There are different RA types for production and ARs as they have different methods to implement, e.g. for the calculation of setup time. The Customer Interface Agent either receives orders through a web service interface, from the database or (for test purposes) generates random orders according to a given distribution. There is also a Database Agent to enter the schedules in a database for further analysis (like GANTT-chart creation) or utilization by an ERP-system, an MES or a simulation.

Before the negotiation scheme and the algorithms are described in detail, a short summary of the architectural motivation from [21] is provided. The general idea of the approach is based on the fact that exact and integrated scheduling of production and transportation processes is facilitated by constraining the complexity of the solution space. This is achieved by scheduling one production stage after another, thus only including one production operation in each scheduling run. In addition, instead of relying on sub-contracting, the OA collects all proposals of each resource type and aggregates the information that need be conveyed to the next resource type. This is also conceptually in line with the findings in [60] where the authors conclude that larger messages pose no significant problem to prominent agent frameworks such as JADE [59].

The communication between the different agents during scheduling is described in the following.

The OA starts the negotiation process by checking the required feature or process, asking for available agents (or using the agents it already knows, depending on the application) at the DF and afterwards contacting the respective production RAs. Depending on the location, time and additional requirements of the proposed production operations, the auxiliary processes are arranged if necessary (more details will be given in Section VII). The negotiation protocol (based on [21] and [11]) follows the well-known Contract Net-Protocol (CNET) by Smith [61] that has been standardized by the FIPA [62]. Among necessary parameters like IDs and deadlines, the following information are included in the messages (the numbers correspond to the numbers in Fig. 3):

1. Call-for-Proposal (CFP): The exact details of the message depend on the recipient's resource type (e.g. production or transport) but in general a CFP contains a desired timeslot (and if applicable the slack before the start and after the end of the timeslot, see sub-section C) as well as details about the workpiece that is asking for an operation and the requested operation itself. Multiple CFPs, created by one OA, can be aggregated within one message and the message can be sent to multiple recipients.
2. Proposal: The proposal message, which is composed in reply to a CFP from one OA, contains proposals on all the CFPs a RA has received. Futhermore, details about the resource are included (like the location), the proposed timeslot and slack (see Section VI), the operation duration, estimated loading times (that can later be adjusted) and the price. The price $PR_{i,n}$ for an operation i on resource n is calculated as follows:
$$PR_{i,n} = T_{Oi,n} + T_{S,Oi,n} + TI_{n+1,n}$$
With $T_{Oi,n}$ being the duration of the operation i on resource n, $T_{S,Oi,n}$ being the setup duration for operation i on resource n and $TI_{i+1,n}$ indicating the time increment (TI) that is required to add or substract from the setup of the operation i+1 of resource n. This change is required as the starting state for operation i+1 is changed to the endstate of the new operation I (see Section VI). Furthermore, the proposal can contain required shared capabilites like those provided by tools or the crane rail or it can also indicate connected operations if, for example, a resource requires a specific follow-up operation.
3. Accept Proposal: The desired and binding timeslot that is booked (referring to the respective CFP and proposal) and the actual unloading time of the chosen transport resource. This message informs the RA that it has won the bidding process.
4. Reject Proposal: Following the speech-act theory that is the basis for the performatives (or types) of the messages [63], the Reject Proposal message represents the act of rejecting a certain proposal. Thus, it contains simply the reference to the correct proposal.
5. Inform Departure: The actual departure time of a workpiece and the actual loading time of the respective transport resource, as well as the reference to the order.

However, to address the different requirements postulated, some adjustments have been made to avoid exponential growth of messages and the indecision problem, as well as incorporate all required functionalities.

Fig. 3 displays the necessary exchange of messages and their sequence, i.e. the protocol, for the coordination of one production stage i. The OA, which usually represents a (large) workpiece but can also represent a composition of smaller workpieces, first contacts the required production resources for the next operation of its process plan (1). For each proposal it then evaluates if SRs, such as operators, are required and contacts these accordingly with the received proposals from the production resources (6). Next it evaluates if buffering is necessary and sends out CFPs accordingly (7). Here it is important to note that the decision upon when a buffering operation is required depends on applicaton-specific parameters that are further explained in Section VI. Following the investigation of possible buffer places, the OA determines necessary transport operations. Here it has, for example, to distinguish between multiple operations (i, i+1) on the same machine and operations on resources that are not connected by a fully automated transportation system like a conveyor belt. Transportation by conveyor belts usually does not require the same level of scheduling granularity as cranes or other flexible resources do.

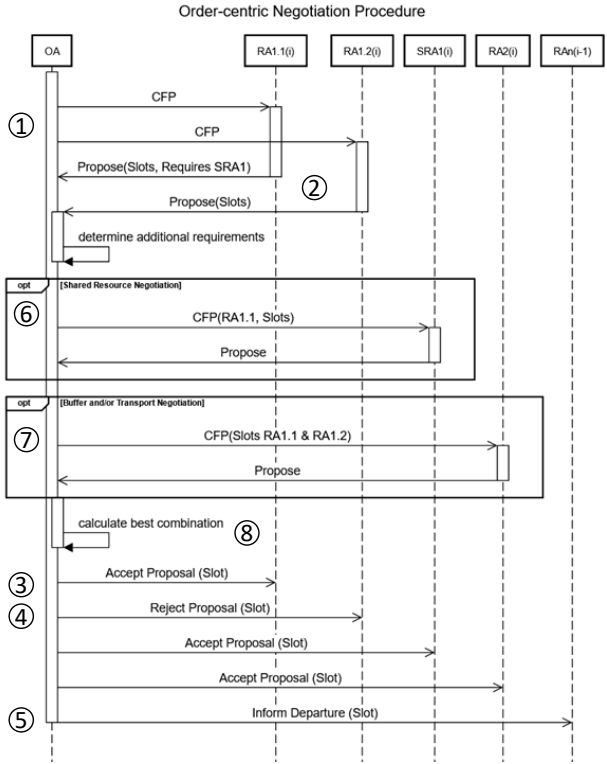

FIGURE 3. Negotiation sequence for one production stage.

The OA needs a transport for all the operations on resources of stage i that have a different location than resource $R_{i-1}$. It therefore combines the proposals by the resources it needs to visit within the CFP which it sends to the transport agent (TA) (7). The TA then calculates a proposal for each requested transport operation, e.g. to a buffer or to a production resource, for every required timeslot. The latter referring to the possibility that a certain operation on a resource can be done at different points in time (see Section VII.A). However, to enable a correct scheduling and optimized resource allocation, the transport proposals need to consider interdependencies between the proposals. Consider the example in Fig. 4 where a workpiece needs to be transported by a transport resource $R_T$ from a resource $R_1$ to either first a buffer place $R_{Buffer}$ and then to resource $R_2$ or directly from $R_1$ to $R_2$. This can, for example, be necessary as $R_1$ can propose two different timeslots of which one starts shortly after the end of the operation that the OA has already planned and the other starts at a later point in time. Hence, t3 in Fig. 4 can be significantly later then t2.

The example displays the resulting reduction of setup times, which results from the combined execution of the two operations by the same resource (P2 and P3). Thus, the TA tells the OA that it can fulfill the transport $R_{Buffer}$ to $R_2$ for a lower price, if it also receives the order for P2. This should be accounted for in the proposals and the Accept Proposal message. Therefore, the proposal P3 indicates a requiredOperation [11] that refers to P1. The accepted proposal also needs to indicate the dependent second operation if applicable.

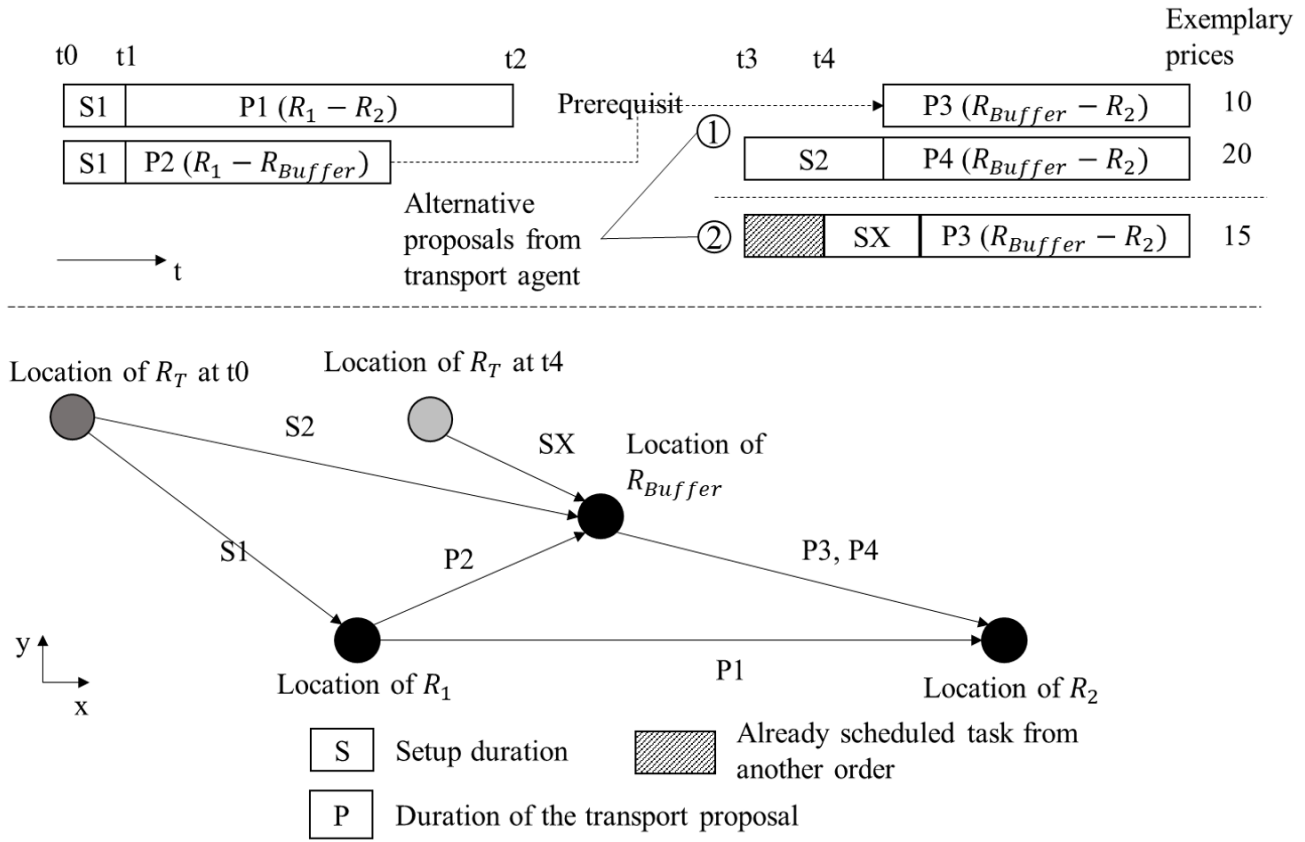

FIGURE 4. Proposal interdependencies.

Furthermore, it is important to indicate the additional setup times a resource has to include in its schedule due to the newly scheduled operation. In correspondence to the general market-based structure of such auction-based coordination approaches, these factors are accounted for in the price of a proposal.

After the OA has received all proposals, it determines the most suitable operation combination (OC) reflecting the applied optimization criterion, e.g. earliest process finish time, and accepts or rejects the proposals (8). The last message in Fig. 3 is important to point out (5) as there needs to be an update to the resource of the production step i-1 regarding the departure time of the workpiece. The devised schedule of each stage i ends upon arrival at a production resource i. Thus, at this time the workpiece's departure from resource I is not yet known and has to be related after the next scheduling run for operation i+1.

This last step is also important for the indecision problem during concurrent scheduling runs by different orders. Only at the moment of receiving and incorporating the Inform Departure message, the RA knows at which exact time it is free again to receive new workpieces. Until that point the RA is not allowed to promise new timeslots to other OAs, i.e. it is blocked for further negotiations. Note that it is also difficult to use estimates at this point to, for example, enable the agent to give away timeslots that are further in the future while locking closer timeslots. This is due to the fact that because of the large workpiece-related constraints it can occur that a production resource is blocked for significantly longer than the actual operation required. This, for example, can happen in case of disturbances or if no buffers are available for storing the workpiece after the current operation and the running process on the next production resource takes longer than the current one. Such an inform departure message can also be sent if the workpiece stays on the machine for another operation to indicate that the next operation on the same resource can start immediately after the other. To avoid deadlocks during the scheduling process in case of individual agent failures or network connection issues, it is necessary to reactivate the agents that are waiting for certain messages after an application-specific timespan has expired. This timespan, for example, can be derived from the estimated time that is required for an average stage to be organized times the number of stages of the application. Another aspect to avoid deadlocks regards the checking procedures of received messages during the auction process. The agent either proceeds if all answers have been received or if a certain deadline has expired. This makes the approach robust against network or software issues that keep agents from answering on CFPs.

The communication with further infrastructure agents that connect the MAS to, for example, a database is ommited in Fig. 4. However, it is worth mentioning that the threading behaviour of such activities can greatly influence the scheduling execution time. Thus, all database related activities should be executed in different threads if the coordination process is required to execute as fast as possible.

Concluding this sub-section, the adherence of the architecture to the formulated requirements from Section II is analyzed.

- Req 1 (Scalability): The presented protocol exhibits a linearly growing number of messages as it does not rely on sub-contracting [11]. The execution time is tackled on a conceptual level by the multi-stage architecture and will also be evaluated in Section VIII.
- Req 2 (Decentralization): The architecture is decentralized without requiring additional hierarchy levels such as coordinator or supervisor agents.
- Req 3 (Dynamic events): As the applied auction mechanism derived from the CNET is inherently capable of including resources during runtime, due to the mechanism of using yellow page services to find suitable agents, Req 3a is fulfilled. Req 3b (disturbance management) is also fulfilled as the scheduling mechanism is also applicable for rescheduling in case of resource failures.
- Req 4 (On-line order generation): Each OA is hosted within the system at the arrival of the respective order. Thus, the scheduling is done on-line.
- Req 5 (Scheduling Scope): The architecture enables the integration of all required functionalities, including production processes and the AR processes as each additional functionality can be scheduled in the process (indicated as optional (opt) in Fig. 3).
- Req 6 (Process Planning): The decentralized process planning [12] is also possible within the architecture.
- Req 7 (Logistics Constraints): The architecture itself does not completely fulfill Req 7 yet (buffer scarcity, blocking constraint and collision avoidance) but provides the basis for the remaining details regarding the required parameters within the messages and algorithms to be applied. Details are provided in Section VI and VII.
- Req 8-11 (Variable Resources and Processes, Variable Operation Sequences, Detailed Auxiliary Process Scheduling, Schedule Stability): These requirements will be addressed by going into more detail in the underlying information model and message exchange in Section VI and the scheduling algorithms in Section VII.
- Req 12 (Concurrency): Req 12 is partly addressed by the architecture by providing the required fundamentals for coping with the concurrency and the indecision problem. The concurrent behaviour of the MAS is also analysed in Section VIII.

## VI. APPLICATION EXAMPLE AND SCHEDULING CONCEPTS

To facilitate the understanding of the function and calculation of different information elements within the coordination process, the following section starts with an industrial application example and further explains the fundamentals, i.e. the required parameters for the scheduling algorithms.

The following example combines the practical requirements of different industrial case studies in the domains of windmill blade manufacturing and aircraft component production. The production facility is a flow shop factory with multiple production, buffer, and transport resources. At the time of the system's initialization, some resources are already partially blocked (looking ahead in their schedule) by orders or maintenance activities. Further orders arrive sequentially (on-line) and contain the following two types of products:

TABLE III.   Product Data

| ID | Product Name | Step | Operation |
|---|---|---|---|
| 1 | A | 1 | Cutting |
|   |   | 2 | Forging |
|   |   | 3 | Milling |
|   |   | 4 | Quality |
| 2 | B | 1 | Cutting |
|   |   | 2 | Forging |
|   |   | 3 | Roll-forming |
|   |   | 4 | Milling |
|   |   | 5 | Quality |

In this case, the production is controlled as an make-to-stock production, meaning that there are no order-specific due dates to adhere to. The optimization criterion is thus for all orders to finish as soon as possible, i.e. to maximize the output by minimizing the throughput time. The following table summarizes the important information of the resources Cutting (C), Forging (F), Roll-forming (Rf), Milling1 (M1), Milling2 (M2) and Quality (Q).

TABLE IV.   Resource Data

| Parameter | C | F | Rf | M1 | M2 | Q |
|---|---|---|---|---|---|---|
| Setup A-B [min] | 15 | 30 | 15 | 15 | 15 | 15 |
| Setup B-A [min] | 30 | 45 | 30 | 30 | 30 | 30 |
| Location (x,y) [m] | 5,5 | 10,11 | 25,15 | 30,5 | 30,15 | 40,5 |
| Op duration [min] | 80 | 150 | 150 | 100 | 100 | 150 |

Note, that the focus of this contribution lies on the integration of production, buffer and transportation scheduling. Thus, the interface-oriented capability modelling and the process planning concepts are simplified whereas the SRs are omitted in this example. Hence, instead of providing detailed capability models based on the Formalized Process Description [11], each resource is capable of fulfilling the operation (Op) that fits to its name. Apart from these production resources there are also two buffer places with a capacity of one workpiece each in the system (at 15,15 and 35,10) and two transport resources (overhead cranes) that move with 1/12 m/s, are able to carry one workpiece and have a load and unload time of ten minutes. The slow speed of the crane has been chosen to facilitate the understanding of the timeslots in the example. Note that the approach does not restrict the speed of transport resources in any way. Each resource can realize its own speed and it does not even have to be constant. A transport operation is structured as follows (based on [64]) using an x-t-diagram that displays the x-coordinate of the crane over time, i.e. its movements during the scheduled operations:

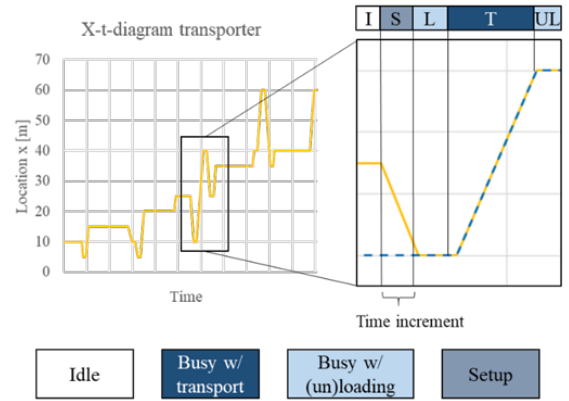

FIGURE 5.   x-t-diagram and transport operation details based on [64].

Each transport operation requries a setup process where the transport resource (in the following "crane") is approaching the loading location. This is marked as time increment (TI) because the required setup time needs to be changed if the preceding operation's end location changes (see previous sub-section). In this case, the full setup is marked as TI but usually the already planned setup time is adjusted by the TI, i.e. its shortened or prolonged by it. When the crane has reached the loading location the workpiece needs to be prepared for transportation. In practice, the workpiece needs to be fixated with straps. This is the loading process during which the production resource is still blocked. Afterwards the workpiece is transported to the next production resource where it is unloaded. Another time increment needs to be incorporated for the next operation scheduled (not marked in Fig. 5). As the focus of this contribution lies not on collision avoidance, the capabilites of the cranes have been set so that each crane serves one, partly overlapping, segment of the factory (Crane 1 [0;30] and Crane 2 [30;60]). This results in the fact that only Crane 1 is capable of fulfilling the first transports in the example and thus proposes a solution (see next section).

The following figure displays a simulation model of the system to illustrate the position of resources and buffers.

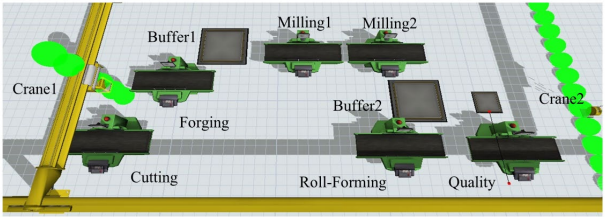

**FIGURE 6.** Discrete-event simulation model of the system.

Because auf the limited knowledge an OA possesses at the beginning of each scheduling run, it first needs to utilize estimations that can later be adjusted with the actual values from the resources capabilites and availabilites. These parameters are the minimal transport time and the minimal buffer duration:

1. Minimal transportation time $T_{T,min}$: The minimal transportation time of the system can be calculated by extracting the minimal distance from the resource table (which results in 5m being the shortest direct pathway in x-direction, assuming that the cranes movement is the restricting factor and not the hoist or trolley [32]) and the maximal speed of transportation resources in the system (being 1/12 m/s). It can therefore be calculated to 2*load/unload time of ten minutes plus one minute travelling time for the shortest distance at maximum speed and results in 21 minutes for the example.

2. Minimal buffer duration $T_{B,min}$: In systems like the one that is presented in the example, a considerable amount of effort is required for the process of buffering a workpiece. The non-negligible transport times need to be incurred twice, including four loading and unloading processes. Thus, the decision should be taken when a dedicated buffering is scheduled compared to a longer stay at the production resource (which delays the arrival of the next workpiece). Hence, $T_{B,min}$ indicates the minimal amount of time a workpiece should remain within a buffer ($B$), i.e. the threshold above which such buffering processes should be scheduled. In case a very small value is chosen, e.g. 1 minute, the transport resources are increasingly busy with the fulfillment of these processes whereas a number that is too large will lead to longer blocking periods of workpieces on resources if, for example, the next production resource is not ready at the right time to enable a smooth transition. The value for this example is chosen at 15 minutes. This value has been chosen because although the transport resources in the system can be the bottleneck, they usually are not due to the long durations of production processes.

Furthermore, some of the other parameters that are important for the scheduling process are also utilized in project planning and scheduling. These include the following (the basic definitions are similar to [65] but vary in detail):

3. Planned Start ($S_{i,o,j}$): The planned start $S_{i,j}$ marks the planned start of an operation i for order j. Note that the index i corresponds to the step number of the OA's process plan. However, in case of buffer operations the number is exchanged with $B$ and for transport operations with $T: i, i+1$, indicating that a transport (T) is planned from required resource for step i to i+1. Due to the Req of plan stability it is not possible to change $S_{i,k}$ for $j \neq k$ (i.e. for already scheduled orders) and $S_{l,j}$ for $l < i$ (i.e. earlier process steps) as they have also been agreed on with the respective resources already). In the following, the index j is omitted when possible as only the scheduling process of the first order is presented. The index o indicates the respective interval to which the planned start refers.

4. Planned Finish ($F_{i,o}$): The planned finish $F_i$ marks the planned finish of an operation i. The same restrictions as for $S_i$ are applicable.

5. Earliest Start ($ES_i$): The earliest point in time at which an operation i can start. All preceding activities must be completed, meaning, for example, that $ES_i$ of the first transportation operation that is required for production step i is equal to $F_{i-1}$ of the preceding production operation. In case of forward looking estimations (if these operations have not been scheduled yet), the estimated minimal time of preceding operations has passed. For example, $ES_i$ of production operation i might depend on the minimal duration of the preceding transport and buffer operations ($T_{T,min}$, $T_{B,min}$). Note that in the latter definition $ES_i$ can be a theoretical value that needs to be aligned with practical availabilites and capabilties of resources. It is serving as a lower bound for the scheduling of further operations.

6. Earliest Finish ($EF_i$): An operation can finish at $ES_i$ plus its operation duration (OpD) or respectively the estimated minimal operation time.

7. Latest Start ($LS_i$): In contrast to the criteria within project scheduling, a start after the $LS_i$ does not delay the overall project but violates the schedule stability constraint. Thus, a later start would lead to a violation of already assigned timeslots to other operations. Note that $LS_i$ of a production operation is constrained by already scheduled maintenance operations or further production operations in the future (of other orders) whereas $LS_{T:i,i+1}$ can be constrained from the production resource I or any of the following resources required for the production operation i+1 (see further explanations in the next section).

8. Latest Finish ($LF_i$): Similarly to the earliest finish the $LF_i$ is calculated by $LS_i + OpD_i$. However, as will be pointed out in the example, $LF_i$ (for example in case of transport operations) is also depending on already scheduled timeslot that demand a specific start date up to which the operation needs to be completed.

9. Slack Time ($ST_{i,o}$): The calculation of STs resembles the critical path analysis but varies in its details. For example, the duration of the different steps is not independent as the duration of buffering depends on the selected slots and exact times of the production and transportation operations. Furthermore, within project scheduling $ST_{i,o}$ represents the difference between $ES_i$ and $LS_i$ of an operation i. In this contribution, the consideration of transportation requires a more detailed approach. As a transport operation includes three separate process steps during which the workpiece is involved (loading, transportation and unloading), the different STs lead to different intervals in which loading, unloading and overall process execution need to take place. The loading process can start ($ES_{T:i,i+1}$) as soon as the production operation has been completed ($F_{i-1}$) and has to be finished before the next workpiece is scheduled for arrival ($S_{i,j-1}$, thus depending on $ST_{i.o,j-1}$). The unloading process at resource i+1 can start when the preceding workpiece k ($j \neq k$) has left production resource i+1 and needs to be finished so that the overall production operation is not finished too late for the next scheduled arrival of a workpiece l ($j \neq l$) or the start of a maintenance operation. In other words, the interval to unload the workpiece j at resource i+1 is determined by the scheduled predecessor operation k on i+1 and the scheduled successor operation l on i+1.

## VII. DETAILED COORDINATION ANALYSIS

The example described in the previous section is now used to illustrate the scheduling process of an order within the system. The necessary details of the interaction are explained that are required for a decentralized detailed schedule generation. Furthermore, the derived schedules are presented and briefly described (sub-section A). Sub-section B presents the algorithm used by the OAs to determine the best combination. This is followed by the description of the algorithms applied by the resources to manage their schedules in sub-section C.

### A. EXEMPLARY SCHEDULING PROCEDURE

The following picture presents the different time intervals that the OA for the first order exchanges with the RAs during the scheduling run for the Forging operation (i.e. the second operation in the OA's process plan). These intervals are communicated in different types of messages, which follow the CNET protocol (see Section V).

The first part of a scheduling run is the negotiation between the OA and the RAs. The following figure shows the intervals and parameters that are relevant during this exchange. The schedule (Sc) of the Cutting RA and the Forging RA as well as the two proposals (P) of the Forging RA are displayed.

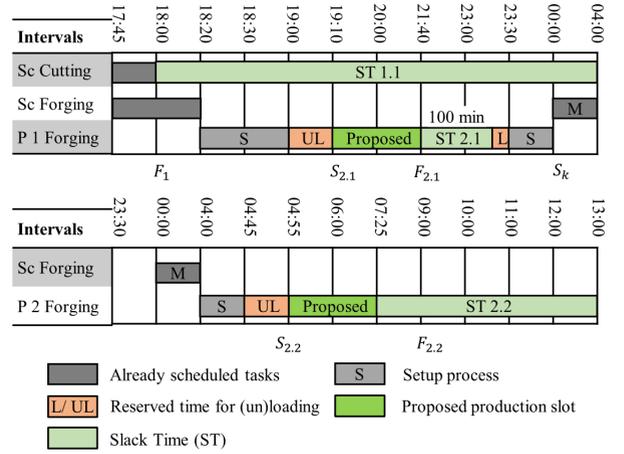

**FIGURE 7.** Production RA and OA negotiation.

The production resource needs to calculate with a setup time (45 min B to A) plus the estimated time for unloading of 10 min, which can later be aligned with the actual value of the transport proposals. ST 2.1 indicates the forging resource's slack time after $F_2$, i.e. time the production operation can start later (100 min). $LS_2$ is at 20:50 so the duration of the production operation (150 min) + loading time (10 min) + setup (45 min) do not violate the planned start of the maintenance task $S_k$ at 0:00.

Following the receival of the production proposal(s), the OA organizes necessary buffers. These buffering operation needs to bridge the time from the finish of the cutting operation $F_1$ at 18:00 and the proposed (and earliest) start $ES_2$ of the forging operation at 19:10. As the exact times of the forging operation are not clear at this point, the buffer place is requested for a longer period of time, including the $ST_{2.1}$. Thus, the workpiece will enter the buffer at 18:00 earliest ($T_{T,min}$ can be neglected here as the buffer will simply propose the slot that fits his schedule). The workpiece must be taken for transportation to Forging within the interval $[EF_B; LF_B]$, with

$$EF_B = S_{2.1} - T_{T,min} \quad (1)$$

$$LF_B = EF_B + ST_{2.1} \quad (2)$$

This results in [18:49; 20:29] for the example. The latest arrival at the buffer $LS_B$ can be calculated from the $LF_B$ minus $T_{B,min}$, i.e. 20:14. These intervals are displayed in the following figure for the first slot proposed by the Forging RA. Naturally, the same mechanisms apply for the second proposed slot.

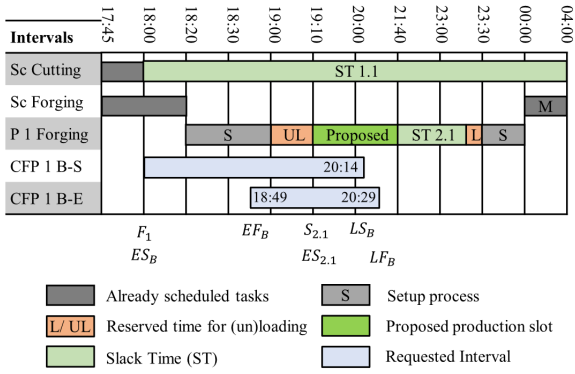

Already scheduled tasks | S Setup process
L/ UL Reserved time for (un)loading | Proposed production slot
Slack Time (ST) | Requested Interval

**FIGURE 8.** Requested buffering intervals.

The buffer agent has no other workpieces scheduled for arrival. Thus, it can propose the requested slot $[ES_B; EF_B]$ and indicate its ST after the slot. Note, that the beginning of the slot must be within the CFP 1 B-S (Buffer-Start) interval and the finish within the CFP1 B-F (Buffer-Finish) interval. If the buffer agent cannot propose such a slot it cannot create a suitable proposal. After the OA has received the buffer proposals, it requests proposals for transportation.

The ES of the transport operation to the buffer place $ES_{T:1,B}$ is at 18:00 when the cutting operation is finished. Note that the indication of the interval number is omitted where possible for simplification purposes. The EF is at 18:00 as well as the buffer place is already available from 18:00 onwards. This is due to the fact that the EF at the buffer place is only restricted by previously buffered workpieces (here again we could add $T_{T,min}$ but it has no effect at this point). The LS needs to be determined as follows:

$$LS_{T:1,B} = min(LF_1, LS_B - T_{T,min}, LS_2 - 2T_{T,min} - T_{B,min}) \quad (3)$$

with

$$LF_1 = F_1 + ST_{1.1} \quad (4)$$
$$LS_B = S_B + ST_{B.1} \quad (5)$$
$$LS_2 = S_2 + ST_{2.1} \quad (6)$$

while ST refers to the slack time after the operations, $F_1$ represents the planned finishing time of the operation 1 and $S_B, S_2$ indicate the planned start times of resources B and i.

The formula shows that the LS depends on the ST after the next scheduled production or buffer operations. This can be extended to further stages of resources if necessary, similar to the traditional methods of the Critical Path Method for calculating the slack or float of activites in a network [65]. $ST_{1.1}$ and $ST_{B.1}$ are large numbers because both resources don't have follow up operations scheduled. Hence, the constraining parameter for $LS_{T:1,B}$ is $LS_{2.1}$ with a proposed start $S_2$ of 19:10 and $ST_{2.1}$ of 100 min, resulting in 20:50 minus 42 minutes for twice $T_{T,min}$ minus $T_{B,min}$. Thus, for the example $LS_{T:1,B}$ is at 19:53. The LF is simply calculated by $LS_{1,B} + T_{T,min}$ resulting in 20:14. Hence, the TA needs to propose a transport that starts within CFP 1 Cutting-Buffer Start (C-B S) and ends within CFP 1 C-B Finish (F).

The transportation operation from the buffer to Forging uses similar parameters for calculation.

$$ES_{T:B,2} = F_1 + T_{T,min} + T_{B,min} \quad (7)$$
$$LS_{T:B,2} = min(LF_B, LS_2 - T_{T,min}) \quad (8)$$

with

$$LF_B = F_B + ST_B \quad (9)$$
$$EF_{T:B,2} = ES_{2.1} = S_{2.1} \quad (10)$$
$$LF_{T:B,2} = S_2 + ST_{2.1} \quad (11)$$

This results in an $ES_{T:B,n}$ of 18:36 and a $LS_{T:B,2}$ of 20:29 and respectively an $EF_{T:B,2}$ of 19:10 and $LF_{T:B,2}$ of 20:50.

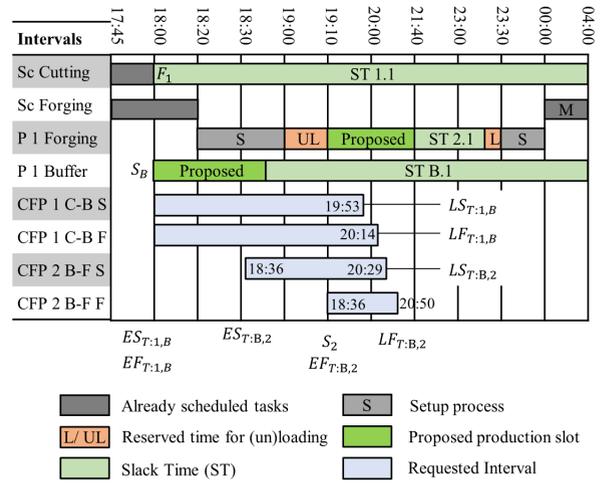

Already scheduled tasks | S Setup process
L/ UL Reserved time for (un)loading | Proposed production slot
Slack Time (ST) | Requested Interval

**FIGURE 9.** Requested transportation intervals.

To fulfill the transportation operations from the buffer to Forging, the TA needs to propose a transport that starts within CFP 2 Buffer-Forging Start (B-F S) and ends within CFP 2 B-F Finish (F). However, the following figure shows that this is not possible for the respective TA in the example.

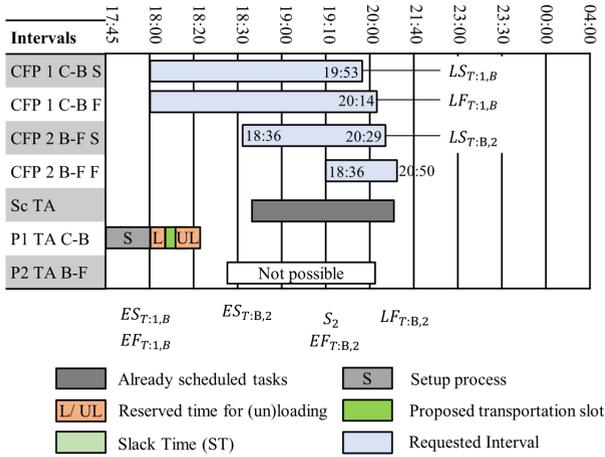

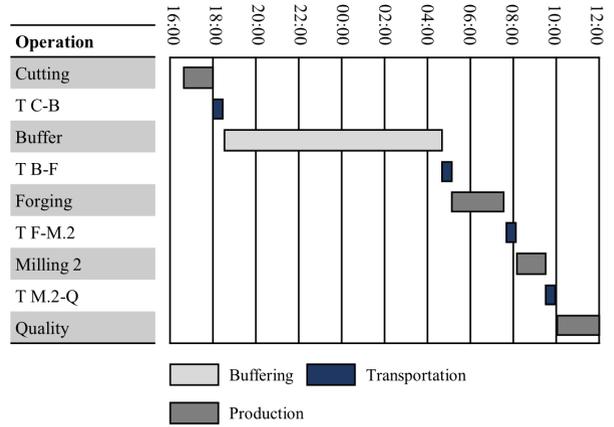

FIGURE 11. Resulting order schedule.

FIGURE 10. Transport proposals for the first production interval.

In Fig. 7 we can see that the second proposal from Forging is much less restricted as ST2.2 is quite large. Thus, there is plenty of time to finish the respective transport operations to the buffer and from the buffer to production. $ES_{T:B,2}$ for this second proposal is equal to $ES_{T:B,2}$ for the first proposal, whereas $EF_{T:B,2}$ is of course different, resulting from the different proposed starting time $S_{2.2}$.

To conclude, the transport proposal includes two slots for two operations. First, there is the transport to the buffer place from 18:00 to 18:22 and later the transport from the buffer place to Forging from 04:34 to 04:55. The transmitted ST before and after each proposed slot can be used by, for example, the crane rail agent that would determine an optimized and collision-free plan (this part of the approach, however, is out of scope of this contribution). As the OA does not receive a proposal for the transportation operation from the buffer to Forging to realize P1 (the earlier proposal), it is forced to accept the proposals that realize P2. In addition to the Accept Proposal messages to the Forging RA, the buffer agent and the TA, an Inform Departure message is sent to the Cutting RA.

The resulting schedule is displayed as a GANTT chart in the following figure.

To avoid confusion the actual position of the workpiece is displayed unambiguously in its schedule. Thus, even though the workpiece blocks more than one resource at a time (during loading and unloading) the operations are shown here as disjunct so that all operation durations added up equal the throughput time of the workpiece. In addition, the workpiece's schedule does not include the resources' setup times. The resource side will be displayed in the production system schedule, which is an aggregation of all the resources' individual schedules. For this, a second order of type B has been launched into the system after the order of type A. The resulting (resource-oriented) schedule is the following:

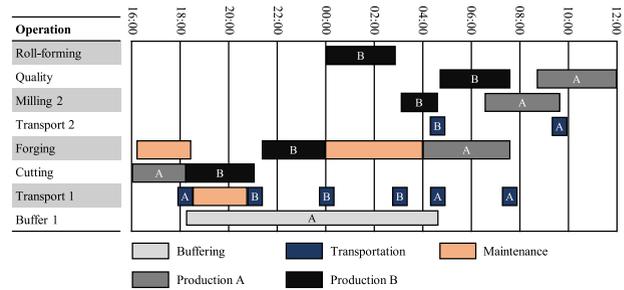

FIGURE 12. Resulting combined resource schedules.

What strikes out is that order B is occupying Forging before A does. This is due to the fact that there is no setup required from B to M (maintenance) in this case and therefore, the MAS can utilize this slot between the already scheduled operations of the transport resource and Forging. Also, the setup times of the resources are visible through the earlier starting times of the slots compared to the arrival of the workpiece, as well as the overlapping occupation of the transport and production resources during loading and unloading.

### B. CALCULATION OF COMBINATIONS
The following sub-section looks at the procedure, how the OA derives the best combination, based on the possible intervals it has received from the RAs and that are presented in the accept proposal part of Fig. 7. The determination of the best possible path is embedded in the OA's control flow as displayed in Fig. 13.

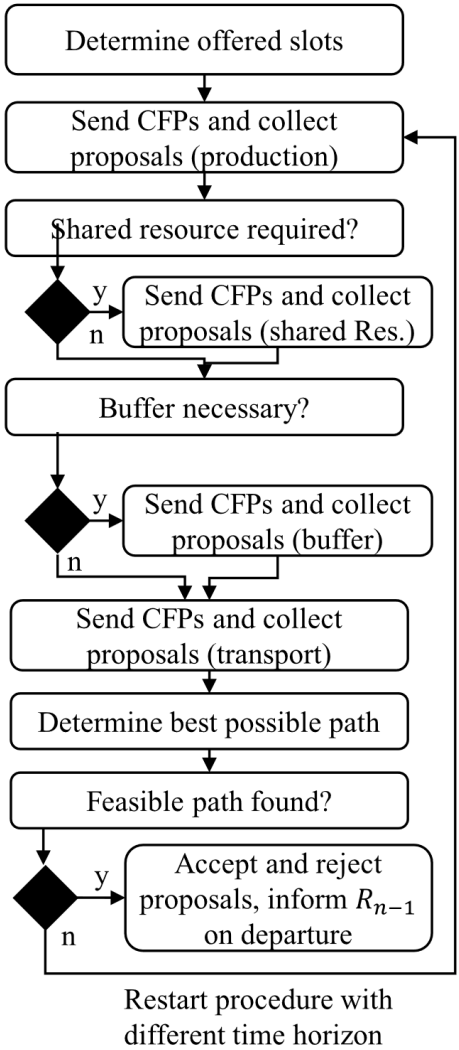

**FIGURE 13.** Order agent's control flow based on [21].

The OA needs to combine different proposals within so-called Operation Combinations (OCs) [21]. Each production proposal that the OA receives constitutes the base for a possible OC. Hence an OC contains all possible routes or possibilities, i.e. different buffers and transport resources, to achieve the fulfillment of this production proposal (P). To determine the best OC, each possible path is calculated and selected according to defined criteria, e.g. earliest finish time or smallest setup duration. This process is displayed in Fig. 14.

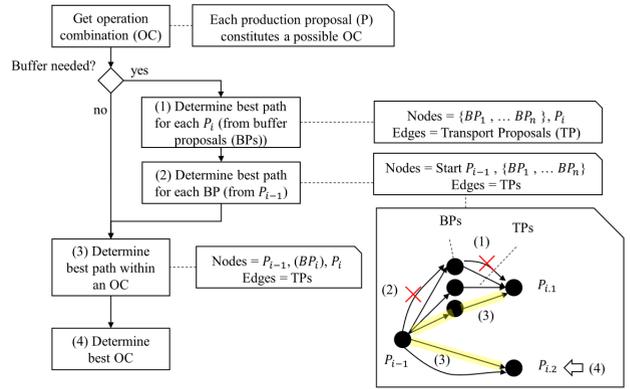

**FIGURE 14.** Calculation of the optimal operation combination.

There are heuristic selection algorithms on multiple levels. Depending on the existence or necessity of buffering the workpiece between the production resources, the calculation algorithm slightly varies. The example in Fig. 14 shows the order of selection processes. In case there are buffer proposals (BPs) required for production proposals $P_i$, the first step is to calculate the best transport operation for each of the different $BP_i - P_i$ combinations, i.e. the best transport solution for a given connection is selected. Second, this step is repeated for each $P_{i-1} - BP_i$ combination, thus resulting in the transport selection for each $P_{i-1} - BP_i$ combination. The red crosses indicate that the best alternative of more than one is chosen, i.e. the one with the red cross is deleted. The third step is to determine the best path for each OC, i.e. the best chain to realize a specific $P_i$. Here, in case of buffering, the different possible routes are weighted against each other, in contrast to the different fulfillments of the same route in the prior step. For OCs without required buffers, i.e. a direct transport $(P_{i-1} - P_i)$, the different transport alternatives are considered in this step. Step three is followed by the selection of the best OC in step four. Note that each selection is based on the agent's local decision-making criteria. These can be the earliest fulfillment, reduced setup times or similar parameters (or combinations of parameters). In the example, the OAs applied the earliest fulfillment date and the price, which includes the setup costs and reductions, as the second criterion in case of equal fulfillment dates.

### C. RESOURCE SCHEDULING PROCEDURE

The resource's behavior during the construction of proposals is already described in [21]. Summarizing that process, the RA receives the CFP and calculates the necessary parameters like duration, which depends on the number of workpieces to produce in that CFP and the desired process, and setup time, which can vary for each slot and depends on the application-specific requirements like possible parallelization of loading and setup processes. It then calculates the possible intervals for that slot and checks their feasibility, incorporating the resource's schedule as well as the information from the CFP. These intervals are based on the different dates the resource receives in the CFP (e.g. ES) and the dates from the RA's schedule. For example, a TA needs to find out if it can start at the ES from the CFP or if it needs to suggest a later start, like the end of its preceding operation. The best possible slot is chosen (again based on local optimization criteria) and the proposal is constructed – including price, (un)loading details and further information. In the example, the RAs propose the earliest fulfillment and incorporate the setup duration in the price.

After the proposal has been sent out, the RA's next step is triggered when an Accept-Proposal message is received. A Reject-Proposal message necessitates no further actions. The following figure displays the process for scheduling a received order.

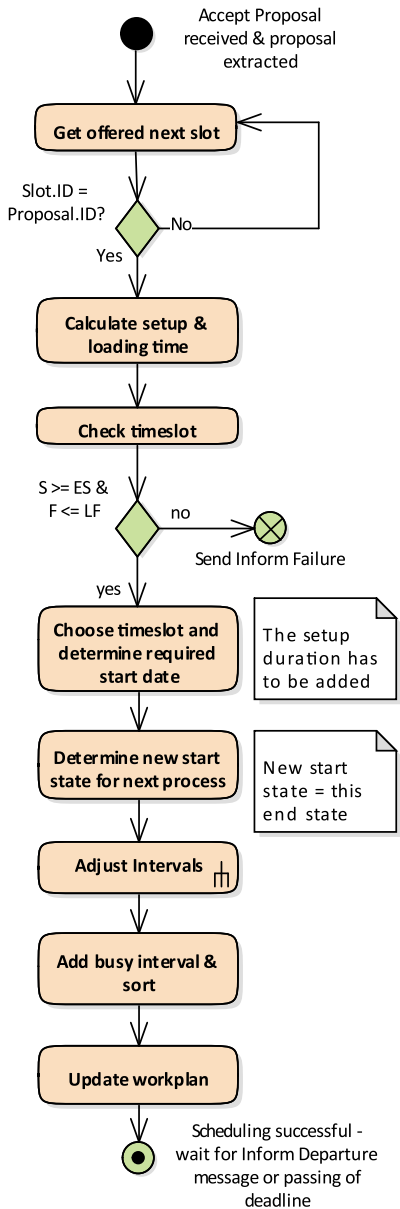

FIGURE 15.  Resource scheduling procedure.

All slots offered from a RA to an OA are stored until either an Accept Proposal message (APM) is received or the deadline for the arrival of Accept or Reject Proposal messages has expired. This is necessary to avoid the pitfalls of the indecision problem, i.e. that slots are promised to more than one OA. The checking of the timeslot determines, if the timeslot contained within the APM does in fact fit to the offered slot and thus fits into the respective free interval. Each RA controls its list of free and busy intervals that, in combination, make up the complete schedule of the agent. The adjustment of the follow-up operation by using the new end state and the required change to the setup process of the next operation (TI) is necessary to avoid errors in the schedule. The free and busy intervals are adjusted accordingly (see next figure), sorted, and the workplan is updated. Note that the workplan contains more information than just the actual intervals of the operation, such as details about the workpiece and the operation (for more details on the workplan see [11]). If the interval within the Accept Proposal message violates the RA's internal scheduling constraints, it sends an Inform Failure message to the OA as the OA has send a wrong timeslot (that was not proposed by the RA).

The adjustments of the intervals follow the logic displayed in Fig. 16:

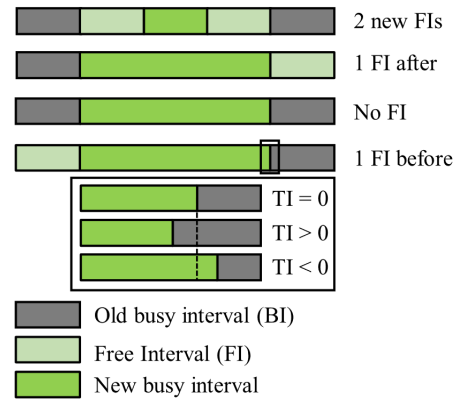

FIGURE 16.  Intervals to be scheduled.

Here again it shows that the TI can also lead to a positive schedule adjustment in the sense that it reduces the setup of the next operation in the schedule.

After this detailed analysis of the coordination mechanisms the following section looks at the performance of the approach in terms of execution time for different scenarios.

## VIII. EVALUATION

To evaluate the approach, a MAS has been implemented in the JAVA-based MAS-framework JADE [59]. JADE provides the agent platform including the DF-, AMS- and Graphical User Interface-agents, basic classes for implementing agent behaviors and communication tools such as the Sniffer-Agent to monitor the agents' communication. The experiments have been conducted on a Windows 7 Notebook with Intel® Core™ i7-3610QM CPU @ 2.30 GHz and 16,0 GB RAM) with all agents running in the same JADE container. Note that a setup with multiple containers on multiple devices can significantly increase the required time to deliver messages [60]. The code implementation described here is freely available on the GitHub agentpro[1] project under the GPL v3.0 license.

Several studies have been conducted to analyze the scheduling execution time and thus scalability. As pointed out in [11] the execution time grows linearly with growing system sizes. The communication architecture ensures that the number of messages does not grow exponentially. Looking at the scheduling execution time per scheduled order in general, there is a slight increase with ongoing order arrival and thus increasing number of orders in the system [21]. This can, on the one hand, be explained by a growing number of intervals that the agents have to check as well as an increasing influence of the JAVA inherent execution time problems such as the garbage collector [66]. The focus of the analysis in this contribution is the behavior of the MAS when many communicative and calculating activities are happening simultaneously.

Generally speaking, an MAS is inherently capable of simultaneous activities and it is one of its core strengths to enable that kind of activity [17, 18]. These can range from different agents engaging in separate negotiations at the same time to different activities within an agent that are executed simultaneously. However, there are important aspects to consider before evaluating this aspect of the approach. First, there is the issue of similarity among the problems the agents have to work on. As described in Section VI, the example is based on an industrial production facility that is organized as a flow shop. If an OA is hosted within such a setup, i.e. it is instantiated with its process plan and the goal to arrange the respective operations, it always needs the resources at the initial positions of the production facility first. Thus, there is increased activity at those first RAs while the rest of the MAS is idle. The first RA becomes a bottleneck if the OAs are hosted in very short succession. The hosting process in the experiments mimics the possible arrival of orders at random points in time in a real production environment. In addition, these hosting intervals can also be applied to regulate the order arrival if more than one order is available to dispatch at the same time. Second, due to the constraints on time and budget for the implementation of this approach there is always the problem of insufficient programming quality. It is well-known that, before a software is used in production, it needs to be developed according to industry standards to avoid unintended behavior during runtime. As the implementation of this work is a proof of concept, it cannot be clearly distinguished how those effects influence the results of the analysis, i.e. how much of the effects need to be attributed to the design of the architecture and algorithms and how much to the implementation and its short-comings. However, it is still a topic of interest to analyze where the boundaries of the approach lie in terms of concurrency.

An experiment has been conducted that analyses different times between the hosting of OAs ($\Delta\ t_{hosting}$) and the resulting differences in start and end times of the coordination process ($\Delta\ t_{start}$ and $\Delta\ t_{end}$). During the experiment the agents are trying to realize a similar process plan as in the experiments in earlier contributions [21]. However, there have been some improvements made in the code of the MAS compared to prior experiments.

First, it needs to be explained when the different times are measured. The starting time of the coordination process is tracked when the OA initializes its production management behavior, i.e. the behavior is taken from the agent's stack of behaviors [59] and actively pursued. This happens right after the OA is initialized. $\Delta\ t_{start}$ reflects the difference of the starting time of the coordination process of $OA_j$ and $OA_{j+1}$. Ideally it should be equal to $\Delta\ t_{hosting}$. The end of the coordination process is measured when all necessary operations of an OA's process plan have been arranged. $\Delta\ t_{end}$ represents the difference between the end of the coordination process of of $OA_j$ and $OA_{j+1}$ accordingly. Note, that the expected ideal behavior of the system would also show an exact correspondence between $\Delta\ t_{hosting}$, $\Delta\ t_{start}$ and $\Delta\ t_{end}$.

The duration of the coordination process displays the difference between start and end of an agent's coordination process. The following figure shows the outcome of the experiment in which 15 agents were hosted for each of the four hosting interval cases.

---

[1] Agentpro https://github.com/FelixGehlhoff/agentpro.git

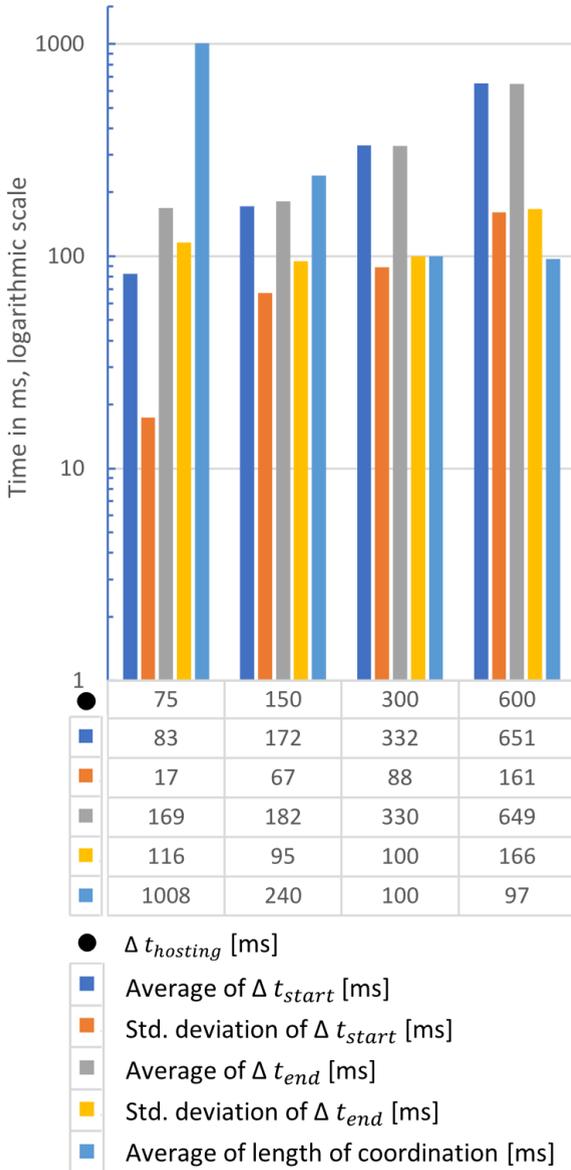

| ● | 75 | 150 | 300 | 600 |
|---|---|---|---|---|
| ■ | 83 | 172 | 332 | 651 |
| ■ | 17 | 67 | 88 | 161 |
| ■ | 169 | 182 | 330 | 649 |
| ■ | 116 | 95 | 100 | 166 |
| ■ | 1008 | 240 | 100 | 97 |

● $\Delta\, t_{hosting}$ [ms]
■ Average of $\Delta\, t_{start}$ [ms]
■ Std. deviation of $\Delta\, t_{start}$ [ms]
■ Average of $\Delta\, t_{end}$ [ms]
■ Std. deviation of $\Delta\, t_{end}$ [ms]
■ Average of length of coordination [ms]

**FIGURE 17.** Execution time evaluation - overview.

The differences between $\Delta\, t_{hosting}$ and $\Delta\, t_{start}$ mainly stem from a few outliers where the start was delayed. This delay can only be explained to inherent uncertainties within JAVA and JADE as there is not much activity within the agent before that point. $\Delta\, t_{end}$ is more interesting to discuss. There is a clear difference between the ideal and resulting behavior of the MAS for the short intervals of 75 ms. Each additional order finishes later than expected, i.e. not 75 ms after the previous order. The length of additional delay for each new order is, however, not increasing overtime but decreasing as Fig. 17 demonstrates. In other words, as long $\Delta\, t_{end}$ stays >0 for sequential pairs of OAs $OA_j$ and $OA_{j+1}$ the resulting delay is adding up.

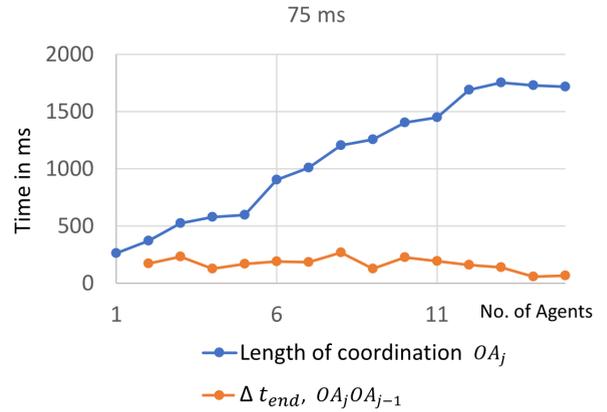

**FIGURE 18.** Concurrency analysis for 75 ms hosting interval.

Fig. 18 shows that the length of coordination for the 75 ms hosting interval is converging at around 1700 ms. Due to this convergence, the additional delay for each order ($\Delta\, t_{end}$) is also converging against 75 ms. To explain this effect, one needs to remember the dependencies between the OAs' coordination processes. Consider the following example: Two orders are hosted in short succession. They require identical machines and thus initially contact RA1. However, to avoid the already described interaction between different proposals (the indecision problem), the receiving behavior of RA 1 is blocked. It can only respond to the second order (OA 2) after it has received the exact departure time from OA 1 (see Fig. 19).

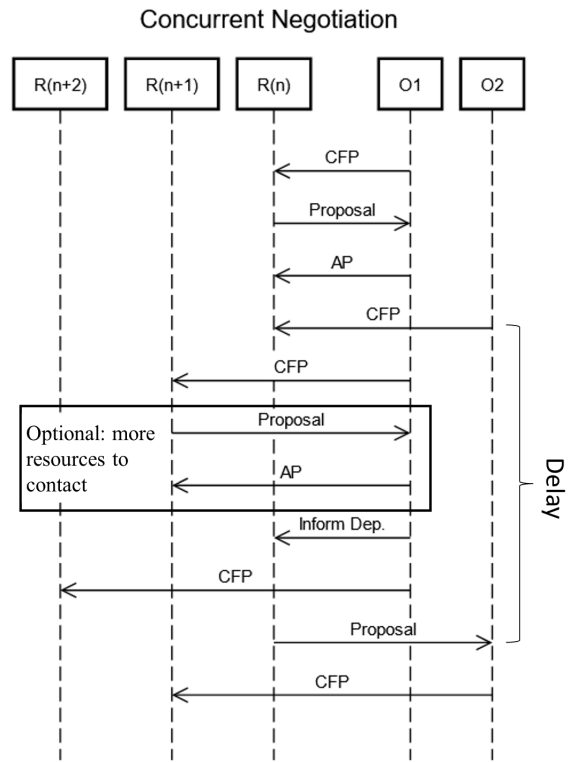

**FIGURE 19.** Protocol-based concurrency analysis.

This dependency and the resulting delays add up as the process continues:
- CFP from OA 3 to RA 1 (step 1) is answered when OA 2 has finished the coordination of step 2 (with RA 2).
- CFP from OA 2 to RA 2 (step 2) is answered when OA 1 has finished the coordination of step 3 (with RA 3).

With the current implementation it is thus not recommended to host multiple orders in such short succession. A possible alternative to longer hosting intervals would be a more sophisticated controlling mechanism of the agent's inbox. For example, as already mentioned, the slots could be on very different parts of the schedule and thus be independent. However, to realize such a differentiation, one would need to consider estimations of latest departure of the workpiece and mechanisms in the software to avoid a so-called "Concurrent Modification Exception" if multiple concurrent behaviors of the agent require access to its schedule simultaneously.

Looking at the other hosting intervals in Fig. 17 it becomes clear that this effect is mitigated strongly by longer hosting intervals. For 150 ms the delay is about 20 percent (average $\Delta t_{end}$ divided by $\Delta t_{hosting}$), for 300 ms it is about 10 percent and even less for 600 ms. The recommended minimum length of hosting intervals for flow shop environments lies at 300 ms, as there are no large improvements for greater numbers.

To analyze the general applicability of the approach, another experiment has been conducted were the coordination time in a flow shop (with product B from Section VI) was compared to a job shop application. The process plans of the orders for the job shop example, which have been generated at random, look as follows:

**TABLE V.** Product Data Job Shop

| ID | Product Name | Step | Operation |
|---|---|---|---|
| 1 | A | 1 | Milling |
|   |   | 2 | Forging |
|   |   | 3 | Cutting |
|   |   | 4 | Roll-forming |
|   |   | 5 | Forging |
| 2 | B | 1 | Cutting |
|   |   | 2 | Quality |
|   |   | 3 | Forging |
|   |   | 4 | Forging |
|   |   | 5 | Quality |
| 3 | C | 1 | Milling |
|   |   | 2 | Forging |
|   |   | 3 | Milling |
|   |   | 4 | Roll-forming |
|   |   | 5 | Milling |

Fig. 20 compares the scheduling execution time in flow and job shop environments.

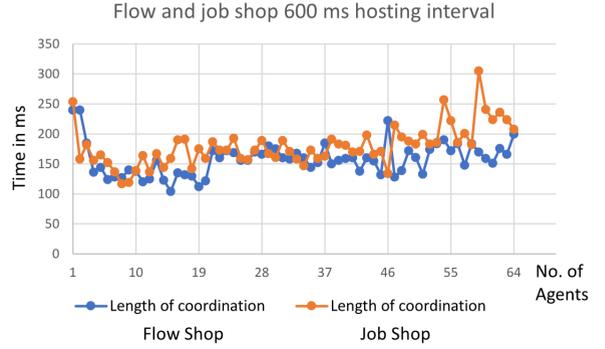

**FIGURE 20.** Evaluation of job- and flow-shop scheduling behavior.

The difference can be attributed to the higher complexity of job shop environments. Due to the randomized order of operations there are more cases in which the MAS is not capable of adding the new operations right next to already scheduled ones. This results in an increased frequency of buffer operations that have to be arranged as the workpieces have to wait for their turn on the next machine more often. Furthermore, the schedules of the agents become more complex as there are less adjacent intervals and thus a higher number of free intervals to work with (see "adjust interval" explanation in the previous section). Hence, the hosting intervals should be chosen with, at least, 300 ms as well.

## IX. CONCLUSION AND OUTLOOK

The advent of ubiquitous information technology and an ever-increasing complexity of production and logistics networks necessitate decentralized solutions to production scheduling and process planning problems. This approach develops a set of parameters to enable the exact scheduling of production and logistics processes while at the same time exhibiting a scalable scheduling execution time behavior in a decentralized architecture. The approach leaves room for individual capabilities of resources and possible extensions like, for example, the incorporation of flexible transportation speeds to enable buffering within transportation resources. The experiments conducted confirm the scalability while the approach provides mechanisms to fulfill the referenced requirements.

However, there are still open issues to address. Further experiments should be conducted to ensure deadlock freedom and collision avoidance for all production and logistics scenarios.

Also, future studies will address the aspect of solution quality in more detail. Even though the optimal solution heavily depends on application-specific parameters such as the buffer threshold, a comparison to a traditional Mixed-Integer Linear Optimization algorithm would lead to further insights on the quality of the scheduling solutions of the MAS. Such an algorithm is currently under development using the open source software OptaPlanner[2]. Early trials show, however, that the scheduling execution time exceeds the proposed threshold of 1-2 seconds considerably.

Furthermore, following the aspect of solution quality, there is also the possibility to improve the scheduling results by incorporating more flexibility. Within the current implementation, the schedule stability is handled very strictly. In further extensions of the approach, there might be the possibility to, for example, change start and end times of an already scheduled operation without changing the sequence of the scheduled orders. Or there can be a certain overlap at the start and end times to account for uncertainty of the operation durations. The strict adherence to the already scheduled plan is currently a clear short-coming of the approach as operations might be added to the far end of the schedule if they don't fit earlier by a margin of seconds.

Another aspect that will be analyzed in the future are scenarios that look at the behavior of the scheduling architecture for large scale production and logistics networks, such as, for example, a network of mid-sized companies connected by free lancing logistics suppliers in a city. Within such a system, the aspects of information hiding and encapsulation are even more important. For instance, a company would not want to relate the details on its already scheduled operations to the customer or any competitor.

**REFERENCES AND FOOTNOTES**

---

[2] https://github.com/kiegroup/optaplanner

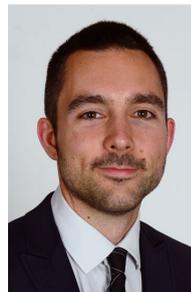

**Felix Gehlhoff** received his B.Sc. degree in Industrial Engineering and Business Management (Wirtschaftsingenieurwesen) in 2014 at the Nordakademie (FH) and his M.Sc. degree in 2017 at the Helmut Schmidt University / University of the Federal Armed Forces Hamburg. He is a senior research associate at the Institute of Automation Technology at Helmut Schmidt University. His main research interests are agent-based systems, especially for production and logistics control.

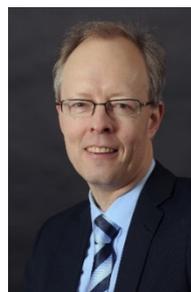

**Alexander Fay** (Senior Member, IEEE) Prof. Dr.-Ing. Alexander Fay (born 1970) is Director of the Institute of Automation Technology at Helmut Schmidt University Hamburg. His main research interests are models, methods, and tools for the efficient engineering of distributed automation systems. Prof. Fay is member of the board of the German association for Measurement and Automation (VDI-/VDE-GMA).